\documentstyle[sprocl]{article}

\input{psfig}

\def\Journal#1#2#3#4{{#1} {\bf #2}, #3 (#4)}

\def\NPB{{\it Nucl. Phys.} B}

\def\PRL{\it Phys. Rev. Lett.}
\def\PRD{{\it Phys. Rev.} D}

\def\be{\begin{equation}}
\def\ee{\end{equation}}
\def\bea{\begin{eqnarray}}
\def\eea{\end{eqnarray}}

\begin{document}

\title{CPT RESULTS FROM KTEV}

\author{HOGAN NGUYEN}

\address{Fermi National Accelerator Lab, \\ Batavia, IL 60510, USA\\
E-mail: hogann@fnal.gov}

\maketitle\abstracts{ I present\footnote{Contribution to the CPT01 Conference, 
Bloomington, Indiana, August 2001.} several preliminary measurements
from KTeV\footnote{Arizona, UCLA, UCSD, Chicago, Colorado, Elmhurst,
Fermilab, Osaka, Rice, Rutgers, Virginia, Wisconsin} 
of the fundamental neutral $K$ parameters, and their implications for
CPT violation.  A new limit is given on the sidereal time dependence of
$\phi_{+-}$.  The results are based on data collected in 
1996-97.}

\section{Introduction}

Neutral kaons have long been recognized as a superb laboratory for the
study of CPT symmetry.  The $K_L$ wavefunction is a coherent 
superposition of a particle and antiparticle:
\begin{equation}
K_L \sim (1+\epsilon_L)K^0 - (1-\epsilon_L)\bar{K^0}
\end{equation}
The powerful probe of CPT symmetry originates from this coherent 
superposition.  There is also a fortunate conspiracy of the kaon masses, 
lifetimes, and decay widths, allowing one to look experimentally 
for very small CPT violating effects that may be present. 
While many probes in the neutral kaon sector are available, I will
mention only ones relevant to KTeV.  Table \ref{observables} shows the 
experimental observables (and their definitions) 
that are sensitive to possible CPT violating effects.

\begin{table}[htpb]
\caption{Experimental observables sensitive to CPT violating
effects. \label{observables}}
\def\Gp{BR(e^+\pi^-)}
\def\Ge{BR(e^-\pi^+)}
\vspace{0.3cm} 
\begin{center}
\begin{tabular}{cc} 
 Observables   &  Definitions \\
\hline
\vspace{.2cm} $\Delta m $, $\tau_S $   &  $M_{K_L} - M_{K_S}$, $K_S$ lifetime \\
\vspace{.1cm} $\phi_{SW}$ &  arctan $\displaystyle{\frac{2\Delta m}{\Gamma_S}}$ \\
\vspace{.1cm} $\eta_{+-}$, $\phi_{+-}$ & $\displaystyle \frac{A(K_L \to \pi^+\pi^-)}
{A(K_S \to \pi^+\pi^-)}$, arg($\eta_{+-}$) \\
\vspace{.1cm} $\eta_{00}$, $\phi_{00}$ & $\displaystyle \frac{A(K_L \to \pi^0\pi^0)}
{A(K_S \to \pi^0\pi^0)}$, arg($\eta_{00}$) \\
\vspace{.1cm} $\epsilon'$  &  $\displaystyle\frac{1}{3}(\eta_{+-} - \eta_{00})$ \\
\vspace{.1cm} $\delta_L$  &   $\displaystyle{\frac{\Ge - \Ge}{\Gp + \Ge}}$ \\
\hline
\end{tabular}
\end{center}
\end{table}

The traditional framework to parameterize possible neutral kaon CPT violation
can be found in the literature \cite{bd,daphnehb,barmin}.  It is a 
phenomological approach: one which spells out all possible CPT effects without
much consideration to the microscopic details of the theory.  
Nevertheless, one can derive the experimental signatures that would 
imply CPT violation. 

CPT violating effects in the $K^0 \to 2\pi$ and 
$\bar{K^0} \to 2\pi$ amplitudes 
can reveal themselves in the $K_L \to 2\pi$ decay mode.  This is possible
since the latter is the {\it coherent~superposition} of those 2 amplitudes.  
CPT violation of this type is called ``direct'' since it is tied to a decay
amplitude.  
Indirect CPT violating effects such as the difference between $M_{K^0}$ and 
$M_{\bar{K^0}}$ could be seen in the comparison of $\phi_{+-}$ to $\phi_{SW}$. 
Direct CPT violating effects may also show up in the semileptonic 
decay modes $K_L \to e^\pm\pi^\mp\nu$ (Ke3).  The two Ke3 final states, 
in the limit of $\Delta S=\Delta Q$ and CPT conservation, would measure 
$2Re~\epsilon_L$, the CP violation in kaon mixing.  
Therefore, a comparison of the semileptonic charge asymmetry
$\delta_L$ to the 2$\pi$ observables might reveal CPT violating effects
in the Ke3 or 2$\pi$ decay amplitudes. 

The following are CPT tests in the traditional framework:
\begin{eqnarray}
\phi_{+-} - \phi_{00} & \sim & \frac{Re B_2}{Re A_2} - 
\frac{Re B_0}{Re A_0} \\
\phi_{+-} - \phi_{SW} & \approx &
\frac{-1}{\sqrt{2}|\eta_{+-}|}\Bigl[\frac{M_{K^0}-M_{\bar{K^0}}}{\Delta m} + 
\frac{Re B_0}{Re A_0}\Bigr] \\
Re(\frac{2}{3}\eta_{+-} + \frac{1}{3}\eta_{00}) - \frac{\delta_L}{2} & = & 
Re~\Bigl( Y + X_{-} + \frac{Re B_0 + iIm A_0}
{Re A_0 + iIm B_0} \Bigr)
\end{eqnarray}
In the equations above, the parameters $B_0$ and $B_2$ ($Y$ and $X_{-}$) 
parameterize direct CPT violating effects in the 2$\pi$ (Ke3) amplitudes.

A significantly new approach called the 'standard-model-extension'
 has been advocated by A. 
Kostelecky and collaborators in recent years \cite{indiana1,indiana2,indiana3}.
It is based on a theory with spontaneously broken CPT/Lorentz symmetry, 
which might happen at the Planck scale in certain string and quantum gravity 
theories.  It appears to be compatible with 
the basic tenets of quantum field theory, and retains the property of
gauge invariage and renormalizability needed for the Standard Model.

The CPT violating phenomenology is significantly 
simpler than the traditional approach.
The direct CPT violating effects vanish at lowest order, so that the lowest
order effect appears in indirect CPT violating observables such as 
$M_{K^0}-M_{\bar{K^0}}$.  An unexpected prediction is a Lorentz violating 
effect such that the 
observables depend on the meson momentum ($\gamma$, $\beta$)
and its orientation in space.  

For a fixed target experiment such as KTeV, one would have:
\begin{eqnarray}
\phi_{+-} & \approx & \phi_{SW} + \\
 &   &   \frac{sin\phi_{SW}}{|\eta_{+-}|\Delta m}
\gamma\bigl[\Delta a_0 + \beta \Delta a_3 cos \chi + 
\beta sin \chi (\Delta a_1 cos\Omega t + \Delta a_2 sin\Omega t)  \bigr] 
\nonumber 
\end{eqnarray}
The angular frequency $\Omega$, time $t$, angle $\chi$ are needed to fully 
describe the orientation of the $K_L$ with respect to $\Delta a_\mu$, which
are the Lorentz violating parameters in this theoretical framework.  For KTeV, 
$\gamma \approx 100$ and cos$\chi \sim 0.6$.  

We present preliminary results from the data taken with the KTeV detector at 
Fermilab during 1996-1997.  It represents half of all data collected by KTeV.

\section{Beam and Detector}

\begin{figure}[tb]
\hspace{2.0cm}
\psfig{figure=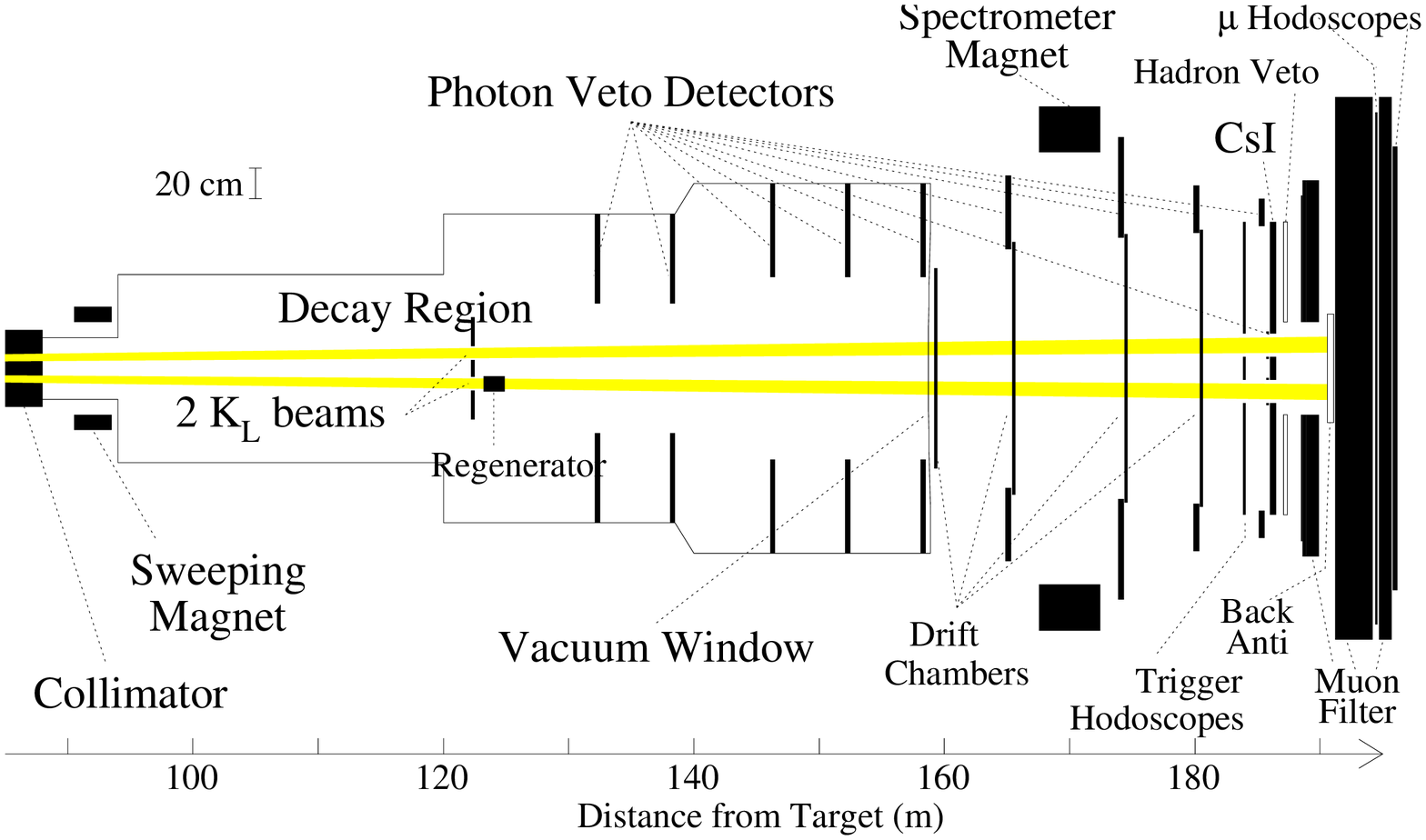,height=3.0in,width=3.5in}
\caption{The KTeV detector configured for measuring
Re($\epsilon'/\epsilon$). \label{detector}}
\end{figure}

The KTeV beamline and detector at Fermilab have been described elsewhere 
in the literature\cite{pss}.  For these preliminary results, 
the detector was configured for the measurement of Re($\epsilon'/\epsilon$).
As shown in Fig. \ref{detector}, two approximately parallel
neutral $K_L$ beams enter a long 
vacuum tank, which defines the fiducial volume for accepted decays.  One 
of the beams strikes an active absorber (regenerator), which serves to tag 
the coherent regeneration of $K_S$.  The regenerator position was moved 
to the other beam in between Tevatron spill cycles.  
Behind the vacuum tank, the charged decay 
products are analyzed by 4 planar drift chambers and an analysis
magnet that imparted a 411 MeV/c horizontal transverse 
kick to the particles.  
A high precision 3100-element pure Cesium Iodide calorimeter (CsI) 
is used primarily to measure the energy of $e^{\pm}$ and photons.  
Photon veto
detectors surrounding the vacuum decay volume, drift chambers, and CsI
serve to reject events with particles escaping the calorimeter.  

\section{2$\pi$ Parameters}

\begin{figure}[tb]
\hspace{2.0cm}
\psfig{figure=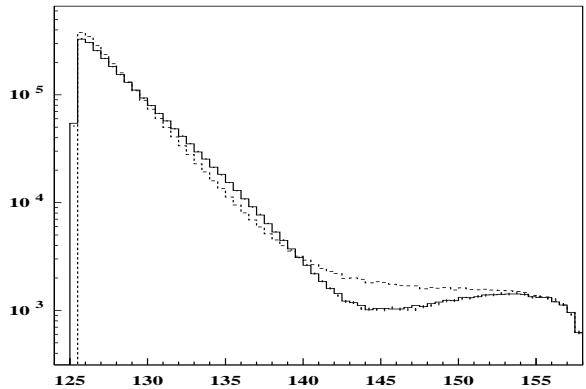,height=2.0in,width=3.0in}
\caption{The 2$\pi$ event rate versus distance (meters) from target for 
$40~<~p_K~<~50$ GeV.
The downstream regenerator edge is located at $\sim$ 125 meters.  
Solid (dashed) histogram shows prediction with (without) interference.
Dots are data. \label{regenerator}}
\end{figure}

The 2$\pi$ parameters $\tau_S$, $\Delta m$, $\phi_{+-}$ and $\phi_{00}$ 
are extracted from data taken with the beam transmitting through the
regenerator.  The regenerator is ``active'', allowing the suppression of 
all but the coherent regeneration of $K_S$. 
With coherent regeneration, the kaon wavefunction is the mixture 
$\rho K_S + K_L$, where $\rho$ is the (momentum and material dependent) 
regeneration amplitude.  Hence the 2$\pi$ parameters can be extracted from
the $K_S$-$K_L$ interference present in the subsequent time evolution of the
kaon wave function. The propertime dependence of the 2$\pi$ event rate 
behind the regenerator is:
\begin{eqnarray}
\label{2pireg}
\frac{dN_{2\pi}}{d\tau} & = & |\frac{\rho}{\eta}|^2e^{-\Gamma_S \tau} + 
 e^{-\Gamma_L \tau} + 
2|\frac{\rho}{\eta}|e^{-\frac{\Gamma_S + \Gamma_L}{2}\tau}
cos(\Delta m + \phi_\rho - \phi_\eta) 
\end{eqnarray}
Figure \ref{regenerator} shows the $K_L-K_S$ interference evident in the
2$\pi$ event rate behind the regenerator.  In practice, one has to subtract 
the incoherent scattering background (inelastic and diffractive), as they 
would have a different time evolution.
\par
These measurements share many common systematics 
with the analysis of Re($\epsilon'/\epsilon$) in the same dataset.  
The latter will be detailed in an upcoming publication.\footnote{The 
preliminary KTeV result from 1996-97 is Re($\epsilon'/\epsilon$) = 
$(20.7 \pm 2.8)\cdot 10^{-4}$.} To determine the event rate
as a function of propertime, one has to reconstruct the parent kaon momentum
$p_K$ and its decay vertex position.  The determination of $p_K$ relies on 
a good understanding of the detector energy scale and its nonlinearities.  
In addition, one has to account for the 
detector geometrical acceptance varying with the decay 
vertex position and $p_K$.

Another necessary ingredient is the understanding of the regeneration 
amplitude $\rho$.  As the data contains inherently only the combination 
$\phi_{\rho}-\phi_{\eta}$, we need a model of $\phi_\rho$ inorder to 
extract $\phi_\eta$.  Our theoretical understanding of $\phi_\rho$ and 
its dependence on $p_K$ includes 
coherent scattering from lead as well as carbon\footnote{The
regenerator contains mostly plastic scintillator (i.e. carbon) and a lead
converter module.}, analyticity, nuclear 
screening, and attenuation.\cite{roybruce,colin}  
In the extraction of $\phi_{\eta_{+-}}$,
we assign a systematic uncertainty of $0.78^\circ$ due to our 
understanding of $\phi_{\rho}$.  We present below a cross check of this in
section \ref{colin}.

The preliminary results are based on the entire 1996-97 KTeV and is
approximately 25M $K \to 2\pi$ events in the regenerator beam.

\subsection{$\Delta m$ and $\tau_S$ \label{deltamtaus}}
\begin{figure}[tb]
\psfig{figure=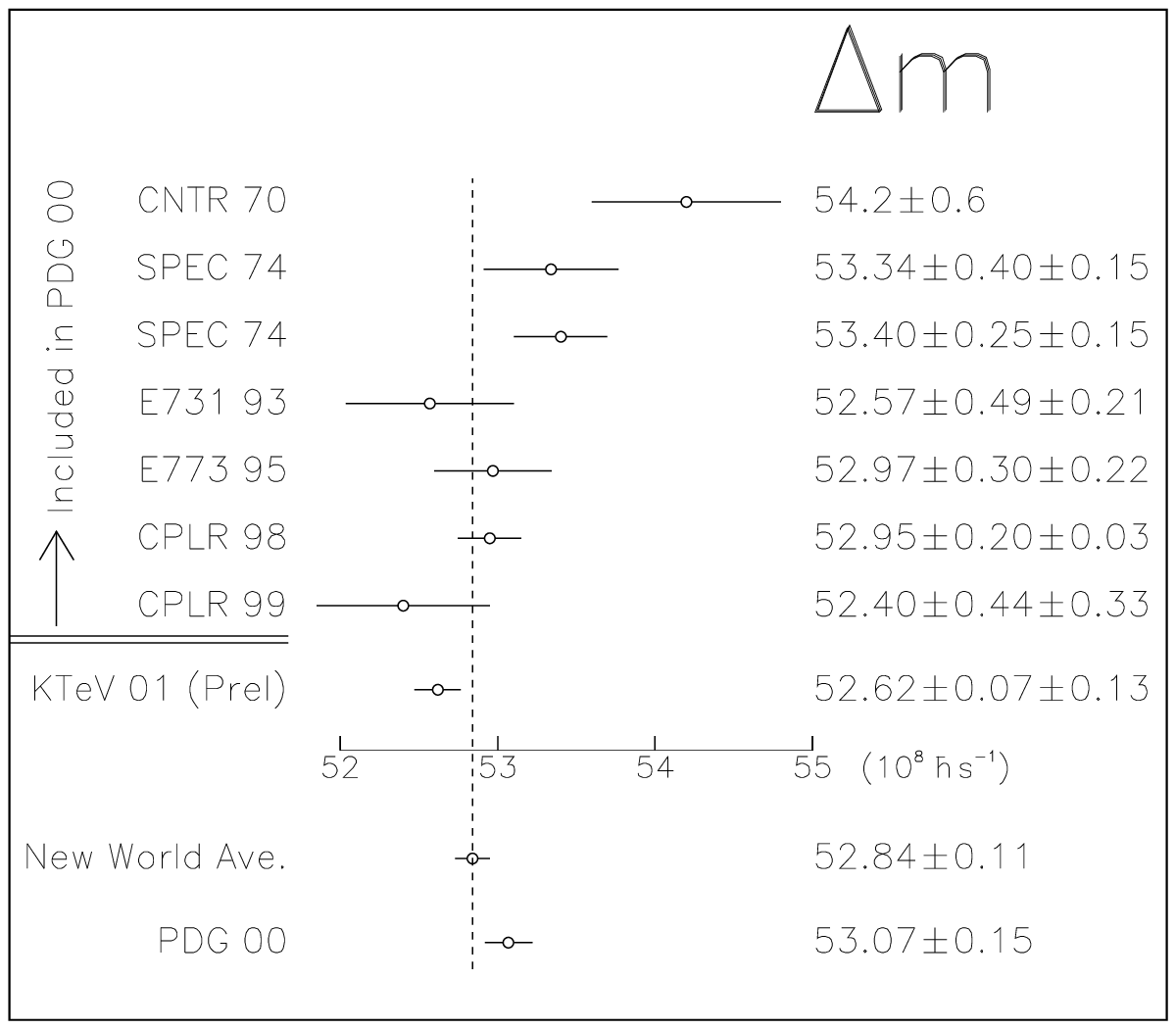,height=2.0in,width=2.0in }
\vspace{-5.1cm}
\hspace{6.0cm}
\psfig{figure=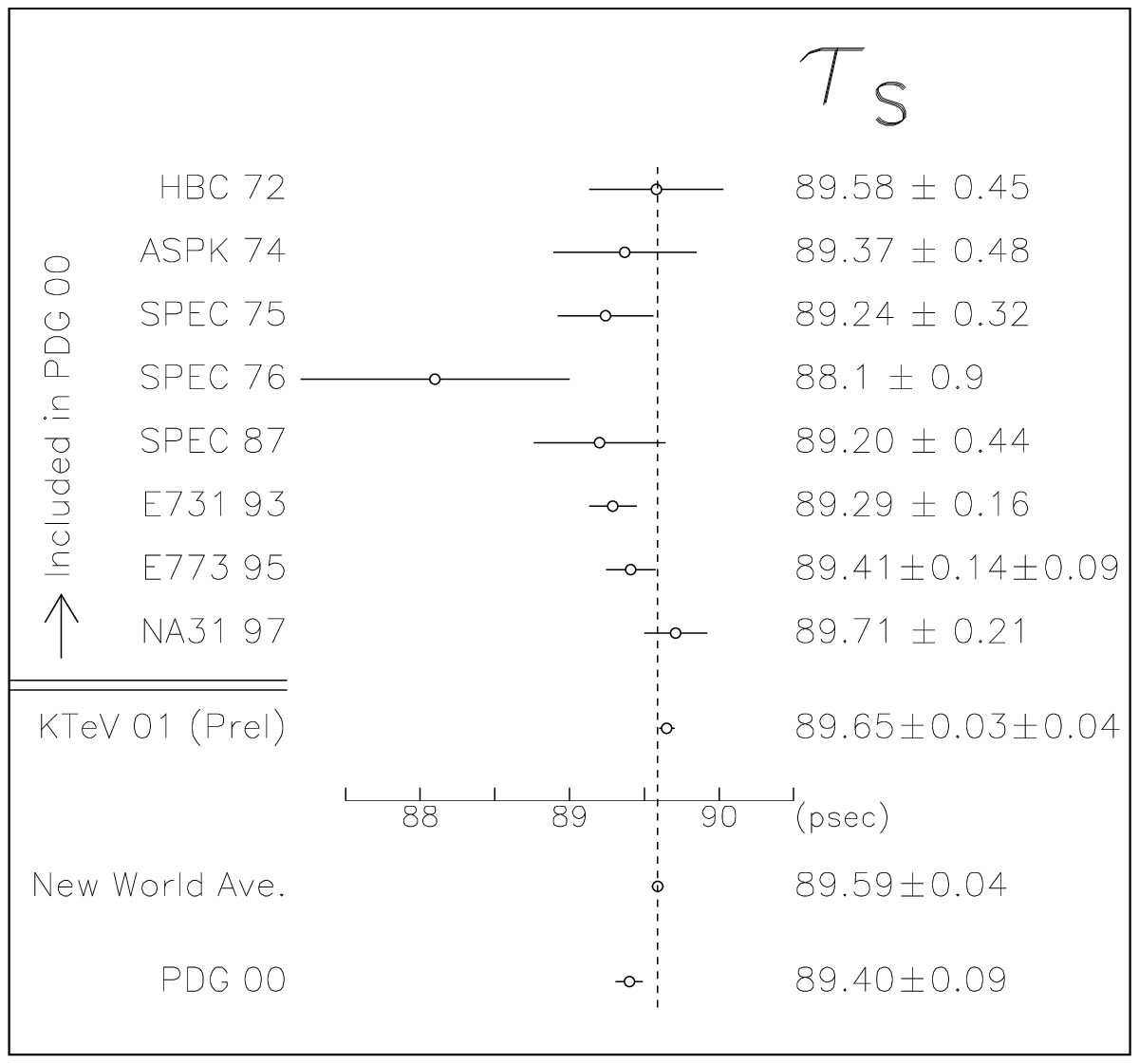,height=2.0in,width=2.0in }
\caption{A compilation of $\Delta m$ and $\tau_S$ values including
our preliminary results. \label{figdeltamtaus}}
\end{figure}

We perform a simultaneous fit of the data
for $\Delta m$ and $\tau_S$, under the assumption of CPT symmetry. That is
to say, we fit with the constraint $\phi_\eta = \phi_{SW}$.  We extract 
a value of:
\begin{eqnarray}
\tau_S & = & 89.67 \pm 0.03_{stat} \pm 0.04_{sys}~{\rm psec} \\
\Delta m & = & 52.62 \pm 0.07_{stat} \pm 0.13_{sys}~10^{8} \hbar s^{-1}
\end{eqnarray}
These are consistent with the PDG\cite{pdg} 
average at only the 2.5 $\sigma$ level.
However, it is more consistent with the published values from the previous 
decade.  Figure \ref{figdeltamtaus} shows a compilation of previous results
including ours.
\subsection{$\phi_{+-} - \phi_{00}$}
To extract $\phi_{+-} - \phi_{00}$, we perform a simultaneous fit of 
the data to all the 2$\pi$ parameters relaxing the constraint of CPT 
symmetry.  However, since the regeneration phase $\rho$ is common 
in both the $\pi^+\pi^-$ and $\pi^0\pi^0$ mode, the phase difference is
immune to the regeneration phase uncertainty.
We find a phase difference that is consistent with no 
direct CPT violation:
\begin{equation}
\phi_{+-} - \phi_{00} = 0.41^\circ \pm 0.22_{stat} ^\circ \pm 0.53_{sys}^\circ
\end{equation}
\subsection{$\phi_{+-}$ and Sidereal Time Dependence}
\begin{figure}[tb]
\psfig{figure=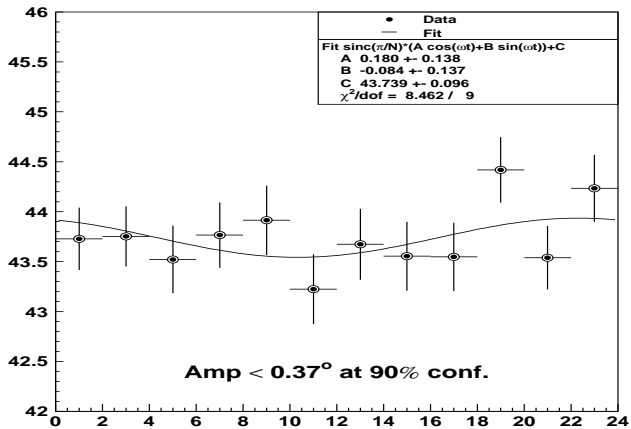,height=2.0in,width=3.0in}
\vspace{-2.0cm}
\caption{The dependence of $\phi_{+-}$ on sidereal time.\label{2pisidereal}}
\end{figure}

The extraction of $\phi_{+-}$ and its comparison to $\phi_{SW}$ is the 
search for indirect CPT violation (i.e. it is sensitive to a nonvanishing
value of $M_{K^0} - M_{\bar{K^0}}$).  We fit simultaneously to 
$\Delta m$, $\tau_S$, and $\phi_{+-}$ with no assumption of CPT 
symmetry, and compare it to $\phi_{SW}$ which was extracted 
under the assumption of CPT symmetry (see section \ref{deltamtaus}).
We find a value consistent with CPT symmetry:
\begin{eqnarray}
\phi_{+-} & = & 44.11^\circ \pm 0.72^\circ_{stat} \pm 1.1^\circ_{sys} \\
\phi_{+-} - \phi_{SW} & = & 0.61^\circ 
\pm 0.62^\circ_{stat} \pm 1.1^\circ_{sys} 
\end{eqnarray}
The systematic 
uncertainties include most notibly the understanding of the 
regeneration phase.
It should be noted that the extracted $\phi_{+-}$ has significant 
correllations with $\tau_S$ and $\Delta m$.  To illustrate the correlation,
we show the variation of the central value of $\phi_{+-}$ in a data subsample:
\begin{eqnarray}
\Delta \phi_{+-} & = & 229.3[(\Delta m \cdot 10^{-10} \hbar^{-1}s) -
0.52796] \\
& & -891.2[(\tau_S \cdot 10^{10} s^{-1}) - 0.89669] \nonumber
\end{eqnarray}
We can now examine the sidereal time dependence of $\phi_{+-}$. 
We perform a fit for only $\phi_{+-}$, while fixing $\Delta m$ and
$\tau_S$ to values extracted ealier (see Eq. \ref{deltamtaus}).  
This would be 
logically inconsistent in the context of the traditional framework.
However, our goal is to test the standard-model-extension, which predicts no
dependence of  $\Delta m$ and $\tau_S$ on sidereal time. 
By similar reasoning, we incur no uncertainty from $\phi_\rho$.  Thus, the
extraction of the sidereal time-dependent amplitudes $\Delta a_{1}$ and 
$\Delta a_{2}$ has nearly only statistical errors.  Figure \ref{2pisidereal} 
shows our result.
We find a 90\% C.L. limit of 0.37 degrees.  Using $cos\chi = 0.6$ and
$\gamma = 100$, we limit $\Delta a_{1}$ and $\Delta a_{2}$ to 
less than $9.2 \cdot 10^{-22}$ GeV at the 90\% C.L.

\section{Charge Asymmetries}

In this section, we present two preliminary results involving the Ke3 charge
asymmetry.  In the beam that does not strike the regenerator, we would measure
$\delta_L$.  In the other beam, we use the charge asymmetry to extract valuable
information about regeneration.  The results are based on approximately 300M
and 150M Ke3 events in the vacuum and regenerator beams respectively.

Since the asymmetries are very small, vary from a few $\%$ to $0.3\%$, we 
combine data from opposite magnet polarity running,
which was reversed approximately once per day.  This ensures that the 
oppositely signed charges have identical acceptance in our detector.

One also has to account for the interaction differences between particle 
and antiparticle in matter (our detector).  
The most significant of these include an asymmetry in the pion energy
deposition in the CsI calorimeter.  There is also a potential asymmetry
in the pion punchthrough rate to our muon counters, which would veto the event
and introduce biases.

\subsection{Regeneration Phase \label{colin}}
We can measure the regeneration phase directly using the charge 
asymmetry in the events behind the regenerator.  Instead of Eq. \ref{2pireg},
the propertime dependence of the charge asymmetry for the regenerated beam is:
\begin{eqnarray}
\delta (\tau) & \approx & (1 + 2Re~x ) \bigl [ 2 |\rho|e^{-\frac{\Gamma_S - \Gamma_L}{2}{\tau}}
cos(\Delta m \tau + \phi_\rho) + \delta_L \bigr]
\end{eqnarray}
The largest uncertainty in the analysis involves the understanding of 
incoherent scatters, which would have a different time dependence and phase.  
Unlike the 2$\pi$ final state, the missing neutrino 
makes it harder to removed scattered background.  Likewise, $p_K$, which is 
needed to reconstruct $\tau$, is known only up to a two-fold ambiguity.  
To deal with the momentum reconstruction uncertainty, 
we rely on our understanding of the detector acceptance.  For the scattered
background, we rely on a model of their time evolution and phase, 
and float their normalization in our fit.
\begin{figure}[tb]
\hspace{2.5cm}
\psfig{figure=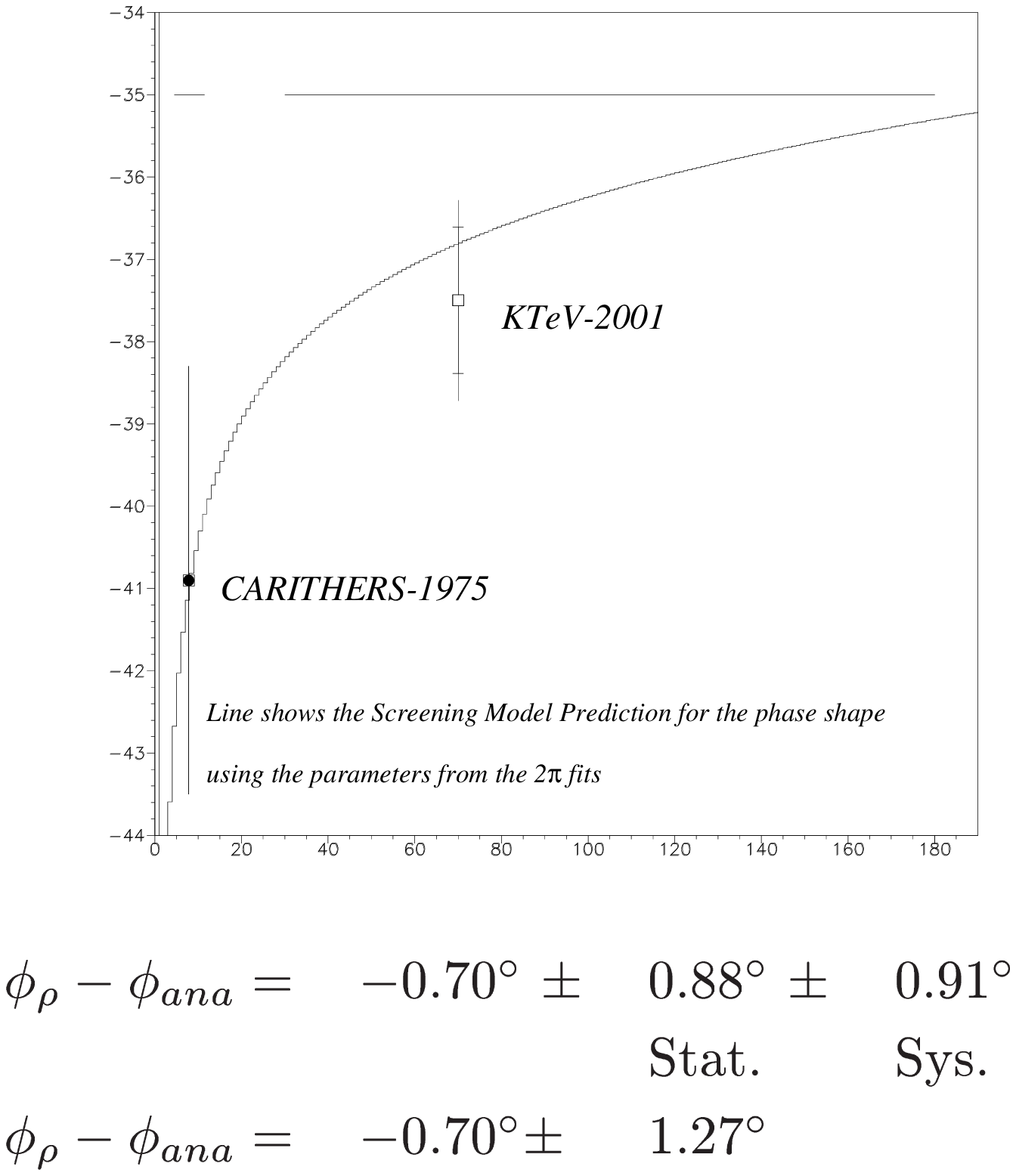,height=3.0in,width=3.0in}
\caption{A comparison of the regeneration phase extracted from the
Ke3 charge asymmetry and our theoretical understanding.\label{analyticity}}
\end{figure}
Figure \ref{analyticity} 
shows our result for the regeneration phase, as well as 
the theoretical prediction.  The agreement of the two is:
\begin{eqnarray}
\phi_\rho - \phi_{th} = -0.70^\circ \pm 0.88_{stat} ^\circ \pm 
0.91_{sys} ^\circ
\end{eqnarray}
We also extract $Re~x$, the real part of the $\Delta S = -\Delta Q$ 
amplitude.  We find a value of:
\begin{eqnarray}
Re~x = (12 \pm 33 \pm 39 )\cdot 10^{-4}
\end{eqnarray}

\subsection{$\delta_L$}
For $\delta_L$, we find a value of:
\begin{eqnarray}
\delta_L & = & 3322 \pm 58~(stat) \pm 47~(syst)~~{\rm ppm} 
\end{eqnarray}
This is in excellent agreement with previous measurements 
and 2.4x more precise than the current best result.  A combination of all 
results including ours yields:
\begin{eqnarray}
\delta_L  = 3307 \pm 63~~{\rm ppm} ~~~~~ 
\chi^2 & = & 4.2/6~~{\rm d.o.f.} 
\end{eqnarray}
The K$\pi$2 amplitude ratios ($\eta_{+-}, \eta_{00}$) 
and $\delta_L$ can be used to place
limits on CPT violation.  Using $\eta_{+-}$ and $\eta_{00}$ values 
from \cite{pdg} and the combined $\delta_L$:
\begin{eqnarray}
\label{cpt}
Re~(Y + \frac{x - \bar{x}}{2} + a) & = & Re(\frac{2}{3}\eta_{+-} + \frac{1}{3}\eta_{00}) - \frac{\delta_L}{2}  \\
 &  = & (1650 \pm 16) - (1653 \pm 32)\nonumber \\
 &  = & -3 \pm 35 ~ {\rm ppm} \nonumber 
\end{eqnarray} 
where $Re~a$ parameterizes CPT violation in K$\pi$2
decays.
The result is consistent with no CPT violation and is limited by the charge
asymmetry uncertainty.  It limits 
$|Re~(Y + (x - \bar{x})/2 + a)| < 61~{\rm ppm}$ at the 90\% C.L..
Barring fortuitous cancellations, this is thus far the most stringent limit on 
$Y,~x,~{\rm and}~a$.

\section{Conclusion}
We have presented several new preliminary results on the fundamental kaon
parameters.  It is based on the entire 1996-97 dataset, which 
represents 1/2 of all data collected by KTeV.  We presented CPT results
in the context of the traditional frame work as well as the one advocated by
Kostelecky and collaborators.  The results are consistent with CPT symmetry.

\end{document}